\documentclass[prb,aps,twocolumn,showpacs,preprintnumbers,superscriptaddress,amsmath,amssymb]{revtex4-1}
\usepackage{graphicx}
\usepackage{dcolumn}
\usepackage{bm}
\usepackage{color}

\begin{document}

\title{Dynamic electronic correlation effects in NbO$_2$ as compared to VO$_2$}

\author{W. H. Brito}
\address{Condensed Matter Physics and Materials Science Department, Brookhaven National Laboratory, Upton, New York 11973, USA.}
\address{Department of Physics and Astronomy, Rutgers University, Piscataway, New Jersey 08854, USA.}
\author{M. C. O. Aguiar}
\address{Departamento de  F\'{\i}sica, Universidade  Federal de Minas Gerais, C. P. 702, 30123-970, Belo Horizonte, MG, Brazil.}
\author{K. Haule}
\address{Department of Physics and Astronomy, Rutgers University, Piscataway, New Jersey 08854, USA.}
\author{G. Kotliar}
\address{Department of Physics and Astronomy, Rutgers University, Piscataway, New Jersey 08854, USA.}
\address{Condensed Matter Physics and Materials Science Department, Brookhaven National Laboratory, Upton, New York 11973, USA.}

\begin{abstract}
In this work we present a comparative investigation of the electronic structures of NbO$_2$ and VO$_2$ obtained within the combination of density functional theory and cluster-dynamical mean field theory calculations. We investigate the role of dynamic electronic correlations on the electronic structure of the metallic and insulating phases of NbO$_2$ and VO$_2$, with focus on the mechanism responsible for the gap opening in the insulating phases. For the rutile metallic phases of both oxides, we obtain that electronic correlations lead to strong renormalization of the $t_{2g}$ subbands, as well as the emergence of incoherent Hubbard subbands, signaling that electronic correlations are also important in the metallic phase of NbO$_2$. Interestingly, we find that nonlocal dynamic correlations do play a role in the gap formation of the (bct) insulating phase of NbO$_2$, by a similar physical mechanism as that recently proposed by us in the case of the (M$_1$) dimerized phase of VO$_2$ (\textit{Phys. Rev. Lett. 
117, 056402 (2016)}). Although the effect of nonlocal dynamic correlations in the gap opening of bct phase is less important than in the (M$_1$ and M$_2$) monoclinic phases of VO$_2$, their presence indicates that the former is not a purely Peierls-type insulator, as it was recently proposed.
\end{abstract}

\maketitle

\section{Introduction}

\begin{figure}[!htb]
\includegraphics[scale=0.38]{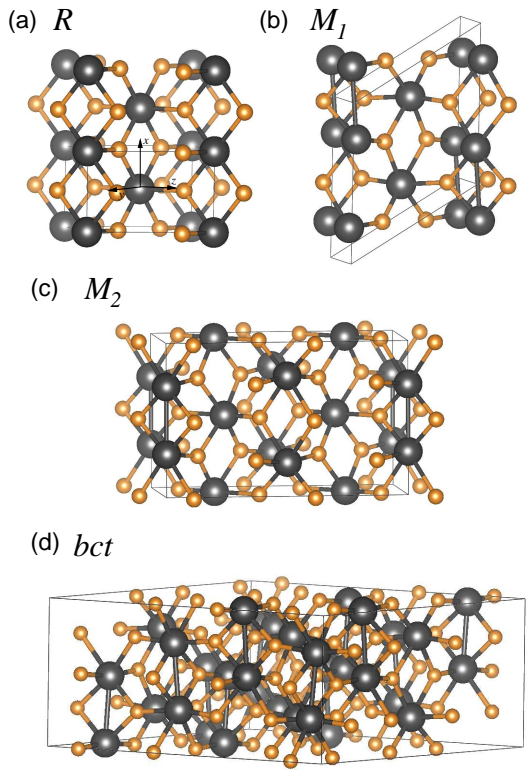}
\caption{Crystal structures of (a) rutile (R) (space group $P4_{2}/mnm$), (b) monoclinic M$_1$ (space group $P2_{1}/c$), and (c) monoclinic M$_2$ (space group $C2/m$) phases of VO$_2$. (d) Crystal structure of body-centered tetragonal (bct) phase of NbO$_2$ (space group $I4_{1}/a$). Vanadium and niobium atoms are represented by black spheres while the oxygens by the orange ones. The local axis system used throughout this work~\cite{eyertAnnPhys} is shown schematically in (a).}
\label{fig:structures}
\end{figure}

Vanadium and niobium dioxides are rutile-based $d^{1}$ systems which undergo simultaneous metal-insulator transition (MIT) and structural transition with dimerization of transition metal atoms.
The MIT in vanadium dioxide (VO$_2$) occurs approximately at 340~K~\cite{morin2} and is accompanied by a transition from a high-temperature rutile (R) phase, shown in Fig.~\ref{fig:structures}(a), to a low-temperature M$_1$ or M$_2$ monoclinic phase (see Fig.~\ref{fig:structures}(b) and (c)).
The MIT in niobium dioxide (NbO$_2$) occurs at much higher temperatures ($\approx$ 1081~K)~\cite{janninck,kseta,ysakai} and is also accompanied by a structural transition, from a rutile to a body-centered tetragonal (bct) phase, displayed in Fig.~\ref{fig:structures}(d). 
From the technological perspective, their ultrafast switching under external stimuli is attractive to engineer new electronic devices. Indeed, both oxides have been considered promising candidates to integrate phase transition electronic devices, such as electronic switches and memristors,~\cite{zyangAnnRev,picketNano,picketNatMat} where the NbO$_2$ has the advantage of operating over a broad range of temperatures.
 
Overall, the M$_1$ and bct structures have similar features. As can be seen in Fig.~\ref{fig:structures}(b) and (d), in these phases the pairs of transition metal atoms dimerize, and tilt with respect to the rutile $c$ axis. Within a band structure picture, Goodenough~\cite{goodenough} proposed that these distortions would lead to the opening of a band gap between the electronic states associated with the overlapping $d$-orbitals along the rutile $c$ axis, namely $a_{1g}$ states, and the remaining $t_{2g}$ states, i.e. the $e_{g}^{\pi}$ states. Thus, within the Goodenough model, the gap in M$_1$ and bct phases opens due to the lattice distortions. 
Meanwhile, the failure of density functional theory (DFT) calculations to take into account the gap opening in the M$_1$ phase,~\cite{wentzprl,eyertAnnPhys} as well as the existence of localized $d$ electrons in the zigzag-like chains of M$_2$ phase,\cite{pouget2} suggest that electronic correlations do play a role in the gap formation of VO$_2$ low-temperature phases.
In fact, according to recent theoretical calculations,~\cite{biermannPRL,cedricPRL,whb} the gap appears in the M$_1$ phase due to the interplay between lattice distortions and electronic correlations, ruling out a purely Peierls-type transition. The calculations performed in Refs.~\onlinecite{biermannPRL,cedricPRL} considered appropriate model Hamiltonians and indicated distinct physical mechanisms for the gap opening of the M$_1$ phase; on the other hand, in Ref.~\onlinecite{whb}, by using modern all electron embedded dynamical mean-field theory (DMFT) implementation, we obtained an unified picture for the gap formation in M$_1$ and M$_2$ monoclinic phases. In particular, we showed that electrons in all phases of VO$_2$ are in the near vicinity of a Mott transition, but with the Mott instability arrested in the dimerized phase.

In NbO$_2$, the role of electronic correlations in the gap formation of bct phase, as well as in the electronic properties of its R phase, has not been addressed so far. Previous DFT calculations, within the local density approximation (LDA), have obtained too small band gap compared to experiment.~\cite{eyertEuro,oharaJAP} Moreover, O'Hara \textit{et al.}~\cite{oharaPRB} have suggested that the structural transition from the R to the bct phase is of second order type in contrast to experiment. They identified soft-mode instabilities in R phase associated with dimerization of niobium atoms. Moreover, they proposed that the gap of bct phase appears solely due to the structural distortions and the electronic correlations play no role. Similar conclusion was reached by Eyert.~\cite{eyertEuro}
Therefore, in contrast to VO$_2$, previous theoretical works have suggested  that the MIT in NbO$_2$ is a structurally-driven transition and that the bct phase is a Peierls-type insulator.

In this work we address the role of electronic correlations on the gap formation of the insulating phases of NbO$_2$ and VO$_2$, as well as in the electronic structure of their respective metallic phases, focusing on a comparison between the two compounds; to address these issues, we use a combination of DFT and embedded cluster-DMFT methods.~\cite{haulepage}
Our results indicate that for the rutile metallic phases the electronic correlations lead to strong renormalization of the $t_{2g}$ subbands and the emergence of incoherent Hubbard subbands in both oxides, signaling that electronic correlations are also important in the metallic phase of NbO$_2$. 
Interestingly, we find the presence of sizable intersite electronic correlations within the niobium dimers in the bct phase. According to our findings, these nonlocal correlations do play a role in the gap formation of NbO$_2$ insulating phase, as found in the M$_1$ phase of VO$_2$.\cite{whb}
Therefore, our results suggest that the structural distortions are not solely responsible for the gap opening of the bct phase, ruling out a purely Peierls-type nature for this phase, as it has been proposed recently.~\cite{oharaPRB}

The paper is organized as follows. In Sec.~\ref{method} we describe the computational method employed in our calculations. Our results for metallic (R) and fully dimerized (M$_1$ and bct) phases are presented in subsections~\ref{res_met} and ~\ref{res_M1bct}, respectively. In subsection~\ref{res_M2} we show our results for the M$_2$ phase of VO$_2$. A general comparison between the nonlocal dynamic correlations in the insulating phases are presented in subsection~\ref{res_nonloc}. Finally, in Sec.~\ref{conclu} we summarize our findings. 

\section{Computational Method}
\label{method}
Our electronic structure calculations were performed within a fully self-consistent combination of DFT and embedded DMFT.~\cite{hauleWK} Within our implementation we do not construct any effective model and the electronic charge density is obtained self-consistently. As shown in our previous report,~\cite{whb} the proper inclusion of ligand states as well as the self-consistent evaluation of the charge density are of great importance to capture the mechanism responsible for the gap opening in VO$_2$. 

In our real space implementation, the DMFT self-energy is expanded in terms of quasi-localized atomic orbitals ($\langle \mathbf{r}|\phi_{m}^{\mu}\rangle$),
\begin{equation}
 \Sigma_{i\omega}(\mathbf{r},\mathbf{r}')=\sum_{mm',\mu\mu'}\langle \mathbf{r}|\phi_{m}^{\mu}\rangle\langle \phi_{m}^{\mu}|\Sigma|\phi_{m'}^{\mu'}\rangle\langle \phi_{m'}^{\mu'}|\mathbf{r}'\rangle,
\end{equation}
where $m$,$m'$ denote the atomic degrees of freedom of an atom centered at $\mu$. The single-site DMFT approximation is obtained by the truncation $\langle \phi_{m}^{\mu}|\Sigma|\phi_{m'}^{\mu'}\rangle = \delta_{\mu,\mu'}\langle \phi_{m}^{\mu}|\Sigma|\phi_{m'}^{\mu}\rangle$, while the cluster-DMFT keeps intersite terms $\langle \phi_{m}^{\mu}|\Sigma|\phi_{m'}^{\mu'}\rangle$ within a given cluster, which in our case are transition metal dimers. In particular, we employed the single-site DMFT for nondimerized atoms and the cluster-DMFT for dimerized ones, such as V-V and Nb-Nb dimers in the low-temperature phases. 
After embedding the self-energy in the large Hilbert space with all valence states included, we solve the Dyson equation
\begin{align}
 G_{i\omega}(\mathbf{r},\mathbf{r}')=[(i\omega+\mu+\nabla^{2}-V_{KS}(\mathbf{r}))\delta(\mathbf{r}-\mathbf{r}')\nonumber \\-\Sigma_{i\omega}(\mathbf{r},\mathbf{r}')]^{-1}.
\end{align}

In our cluster-DMFT treatment, we adopt the symmetric and antisymmetric combination of orbitals within each transition metal dimer. The associated bonding ($\Sigma_{b,\alpha}$) and antibonding ($\Sigma_{ab,\alpha}$) self-energies can be expressed as a linear combination of the components in the site representation, i.e. in terms of a local ($\Sigma_{11}$) and intersite ($\Sigma_{12}$) components,
\begin{equation}
 \Sigma_{b,\alpha} = \Sigma_{11}+\Sigma_{12},
\end{equation}
and
\begin{equation}
 \Sigma_{ab,\alpha} = \Sigma_{11}-\Sigma_{12},
\end{equation}
where $\alpha=\{a_{1g},e_{g}^{\pi}(1),e_{g}^{\pi}(2)\}$.

Finally, we mention that the DFT part of our calculations were carried out within Perdew-Burke-Ernzehof generalized gradient approximation (PBE-GGA),~\cite{pbe} as implemented in Wien2K package.~\cite{wien}
In our DMFT calculations, the quantum impurity problem was solved using Continuous time quantum Monte Carlo (CTQMC) calculations,~\cite{ctqmc} considering the Coulomb interaction $U = 6.0 $ eV and Hund's coupling $J = 1.0$ eV for all phases investigated. In our notation, the $a_{1g}$ state corresponds to the $\sigma$-type $d_{x^2-y^2}-d_{x^2-y^2}$ overlap along the $c$ axis, while the $e_{g}^{\pi}$ states correspond to the $\pi$-type $d-d$ overlap concerning the $d_{xz}$ and $d_{yz}$ orbitals. The experimental lattice structures of VO$_2$ and NbO$_2$ phases are taken from Refs.~\onlinecite{mcwhan}, \onlinecite{jmlong}, \onlinecite{marezio}, and \onlinecite{bolzan}, respectively.

\section{Results and Discussions}

\subsection{Metallic phases}
\label{res_met}

\begin{figure}[!htb]
\includegraphics[scale=0.35]{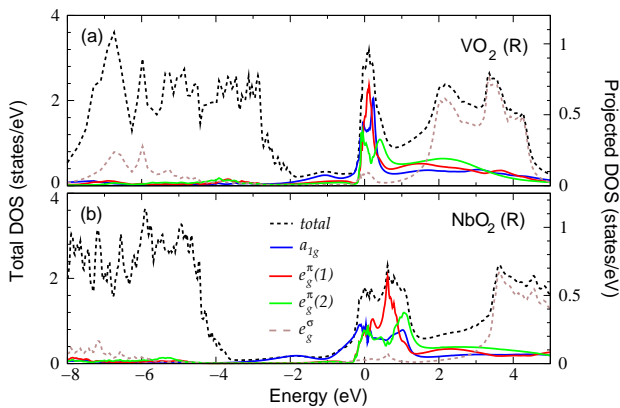}
 \caption{DFT+DMFT-based total (black dashed line) and projected density of states of R  phases of (a) VO$_2$ and (b) NbO$_2$, at $T = 390$~K and $T = 1132$~K respectively. The projections to $a_{1g}$, $e_{g}^{\pi}(1)$, $e_{g}^{\pi}(2)$, and $e_{g}^{\sigma}$ states are shown in blue, red, green, and brown lines, respectively.}
\label{fig:dos_R_vnbo2}
\end{figure}

We first investigate the R phase of both oxides within our realistic DFT+DMFT approximation. In Fig.~\ref{fig:dos_R_vnbo2} we show the calculated DFT+DMFT based total, $t_{2g}$, and $e_{g}^{\sigma}$ projected density of states of the rutile phase of both oxides, at temperatures close to their respective MITs.

In comparison with DFT obtained density of states (not shown), we notice that the $t_{2g}$ states are renormalized due to dynamic correlations, with stronger renormalization in VO$_2$. As can be seen in Fig.~\ref{fig:dos_R_vnbo2}, these correlations also lead to the emergence of lower and upper Hubbard bands in the spectra of both oxides. For VO$_2$, a lower Hubbard band (LHB) is found at -1.1 eV, in agreement with experimental measurements~\cite{koethePRL} and DMFT calculations on the Hubbard model.~\cite{biermannPRL} The upper Hubbard band (UHB), by its turn, is observed at around 2.5 eV. For NbO$_2$ the LHB is found at $\approx$ -1.8 eV whereas the UHB at $\approx$ 3 eV. To date, there is not any experimental spectra reported on the R phase of NbO$_2$, which could be used for comparison with our findings. Overall, the LHBs come mainly from $a_{1g}$ states, while the UHBs are mainly due to $e_{g}^{\pi}$ states. Although the correlation effects are more pronounced in spectral properties of VO$_2$, our 
findings suggest that the electronic 
dynamic correlations in NbO$_2$ are still important.

To investigate the strength of correlations in both oxides, we evaluated the quasiparticle weight $Z_{\alpha} = [1-\partial \Re\Sigma_{\alpha}(\omega)/\partial\omega]^{-1}$ for each dynamical orbital $\alpha = \{a_{1g},e_{g}^{\pi}(1),e_{g}^{\pi}(2)\}$. For non-interacting systems, this quasiparticle weight is equal to unity, while in a strongly correlated system, such as a Mott insulator, $Z$ vanishes. Our calculated quasiparticle weights for both oxides are presented in table~I. 

\begin{table}[!htb]
\label{rutiles_res}
\caption{Quasiparticle weights (Z's) for each dynamical orbital $\alpha = \{a_{1g},e_{g}^{\pi}(1),e_{g}^{\pi}(2)\}$ of rutile phases of VO$_2$ and NbO$_2$.}
\begin{ruledtabular}
\begin{tabular}{ccc}
   & VO$_2$ & NbO$_2$                        \\ \hline 
$Z_{a_{1g}}$         &  0.28  &  0.32         \\   
$Z_{e_{g}^{\pi}(1)}$&  0.33  &   0.51           \\ 
$Z_{e_{g}^{\pi}(2)}$ &  0.40 &   0.57            \\
$Z_{avg}$  & 0.34  &  0.46                  \\
\end{tabular}
\end{ruledtabular}
\end{table}

The obtained values of $Z$ for each $t_{2g}$ state indicate that the metallic phase of VO$_2$ is indeed more correlated than that of NbO$_2$. However, we stress that the $Z$ values obtained for NbO$_2$ confirm that correlations are also important in this system. In particular, we observe that the $a_{1g}$ subband is the most correlated, followed by the $e_{g}^{\pi}(1)$ and $e_{g}^{\pi}(2)$ subbands. The smaller values of $Z$ obtained for VO$_2$ reveal that electrons in this system are closer to the Mott transition than electrons in NbO$_2$, which is in accordance with the more delocalized nature of $4d$ orbitals of niobium in comparison with the $3d$ ones of vanadium atoms. 

\subsection{Insulating phases}

\subsubsection{M$_1$ and bct phases}
\label{res_M1bct}

Structurally, in the M$_1$ phase of VO$_2$ as well as in the bct phase of NbO$_2$ the transition metal atoms dimerize and tilt with respect to the rutile $c$ axis, as shown in Figs.~\ref{fig:structures}(b) and (d), respectively. These structural distortions, within a band-theory, lead to the splitting of the $a_{1g}$ subband in bonding and antibonding states, while the $e_{g}^{\pi}$ states are upshifted in comparison with its respective position in rutile.
For the M$_1$ phase, previous DFT+DMFT calculations showed that the gap opens due to the interplay of structural distortions and electronic correlations.~\cite{biermannPRL,cedricPRL} However, their findings pointed out different mechanisms for the gap formation. In fact, in the work by Biermann~\textit{et al.}~\cite{biermannPRL} it was found that non-local correlations, within the vanadium dimers, renormalize down the $a_{1g}$ bonding-antibonding splitting, suggesting that the M$_1$ phase should be a renormalized Peierls insulator. In contrast, Weber~\textit{et al.}~\cite{cedricPRL} found that the gap formation is driven by an orbital-selective Mott instability of the $a_{1g}$ electronic states, suggesting that the M$_1$ should be a Mott-Peierls insulator. More recently, we showed~\cite{whb} that the gap appears in the M$_1$ due to significant non-local correlations in the presence of strong intersite exchange within the vanadium dimers; this rules out the previous findings in favor of a Mott insulator in 
the presence of strong intersite superexchange within V-dimers. 

\begin{figure}[!htb]
 \includegraphics[scale=0.67]{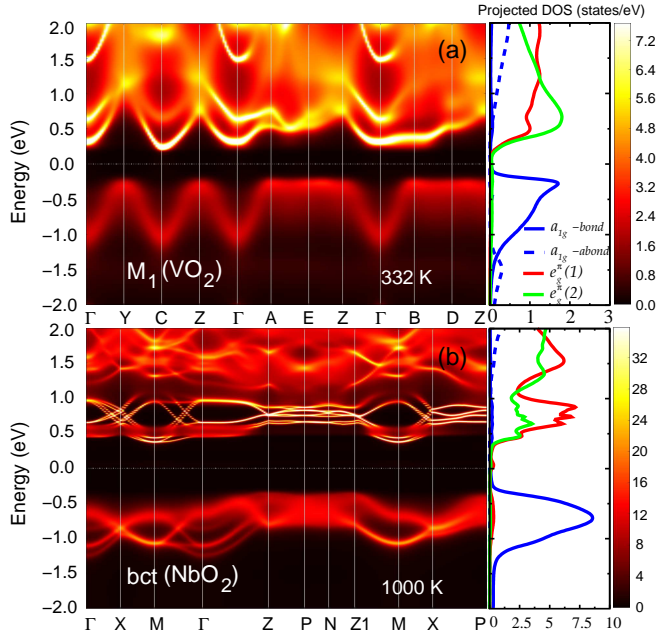}
\caption{Spectral function and projected density of states of (a) M$_1$ and (b) bct phases at 332 and 1000 K, respectively. The projections to $a_{1g}$, $e_{g}^{\pi}(1)$, and $e_{g}^{\pi}(2)$ dimer states are shown in blue, red, and green lines, respectively. The solid (dashed) blue line corresponds to the projection on the bonding (antibonding) $a_{1g}$ dimer state.}
\label{fig:akw_M1bct}
\end{figure}

To compare the low temperature phases of VO$_2$ and NbO$_2$, we show in Fig.~\ref{fig:akw_M1bct} the calculated spectral function of M$_1$ and bct phases, with their associated $t_{2g}$ and $e_{g}^{\sigma}$ projected density of states.
Within our DFT+DMFT approach, we obtained two insulating phases with indirect gaps of 0.55 and 0.73 eV, for M$_1$ and bct phases, respectively. 
As mentioned in Ref.~\onlinecite{whb}, the gap obtained for M$_1$ is in good agreement with the experimental gap reported by Koethe et al.~\cite{koethePRL} (0.6 eV) and cluster-DMFT calculations on the Hubbard model.\cite{biermannPRL,cedricPRL} In bct phase of NbO$_2$, we obtained the indirect gap size of 0.73 eV, which is 0.49 eV higher than DFT(GGA) band gap and is in good agreement with recent ellipsometric measurements reported by O'Hara~\textit{et al.}~\cite{oharaJAP} (gap of 0.7 eV). It is worth mentioning, though, that our charge gap is underestimated in comparison with the gap of at least 1.0 eV obtained by Posadas~\textit{et al.}~\cite{posadasAPL} through x-ray photoelectron spectroscopic measurements of NbO$_2$ films.
We notice also that, in both oxides, the $e_{g}^{\pi}$ states are more coherent than the $a_{1g}$ states, which suggests that the latter is more correlated than the former. Further, the weak LHB associated with the $a_{1g}$ antibonding state in M$_1$ phase of VO$_2$ is not seen in bct phase. The $a_{1g}$ bonding subbands present coherent peaks at around -0.3 and -0.7 eV in M$_1$ of VO$_2$ and bct phases, respectively. The respective antibonding subbands are centered around 2.6 and 3.5 eV.
This indicates that the bonding-antibonding splitting energy has an increase of 1.34 eV in M$_1$ phase and 0.74 eV in bct phase, in comparison with our DFT calculations.

\begin{figure}[!htb]
 \includegraphics[scale=0.47]{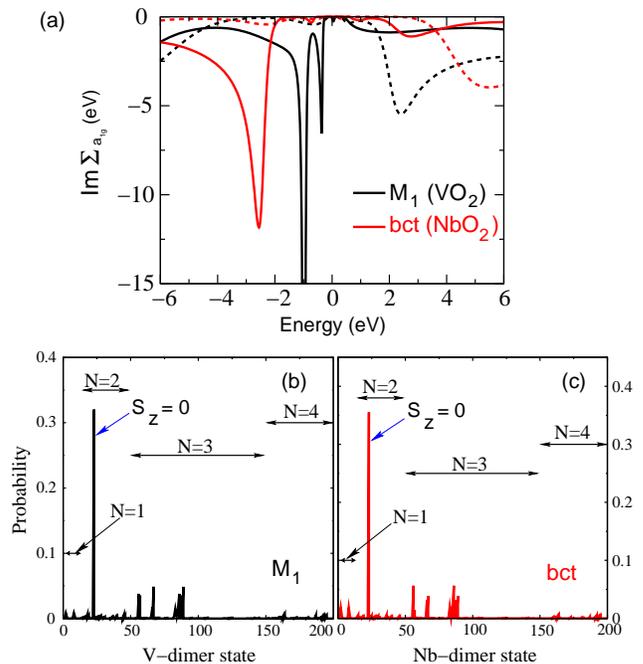}
\caption{(a) Imaginary part, on real frequency axis, of bonding (dashed lines) and antibonding (solid lines) self-energies associated with the $a_{1g}$ dimer electronic states of M$_1$ (black) and bct (red) phases. In (b) and (c) we show the valence histograms of vanadium and niobium dimers in M$_1$ and bct phases, respectively. In these figures $N$ denotes the number of electrons in distinct dimer states. Singlet states ($S_z = 0$) are indicated by blue arrows.}
\label{fig:self_M1bct_onreal}
\end{figure}

Next, in Fig.~\ref{fig:self_M1bct_onreal}(a), we show the imaginary part of the self-energies associated with the $a_{1g}$ dimer electronic states, for both M$_1$ (black) and bct (red) phases. 

\begin{figure*}[!htb]
\includegraphics[scale=0.7]{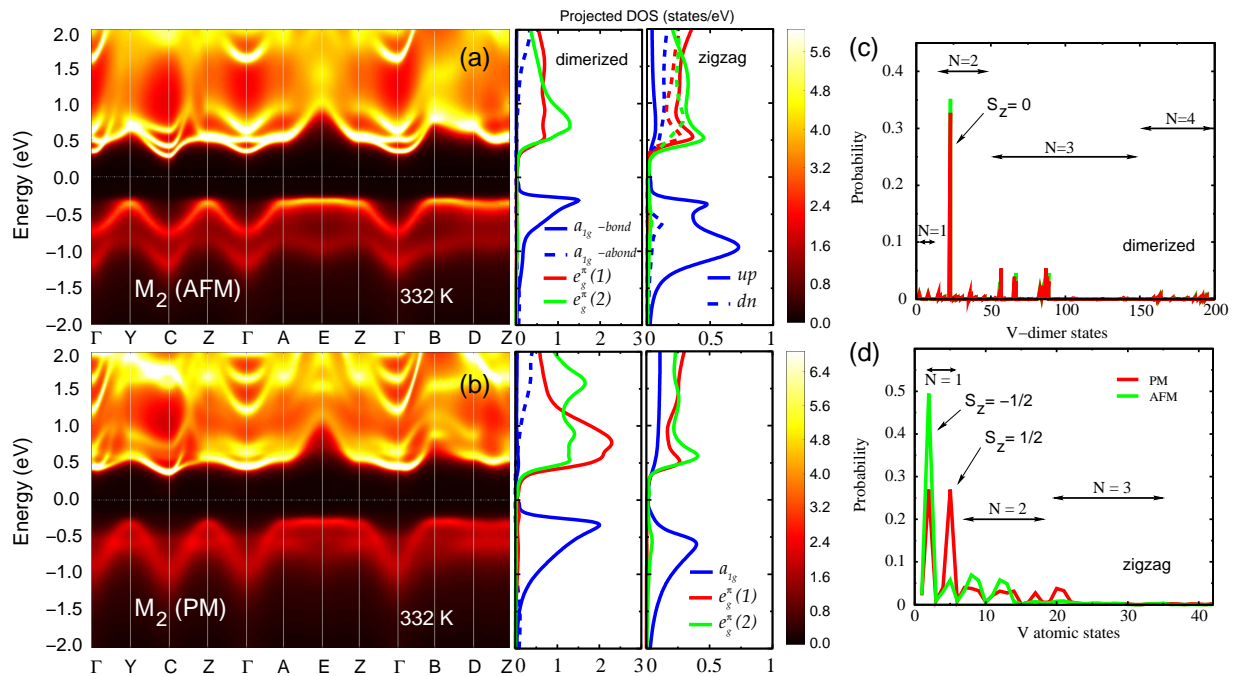}
\caption{DFT+DMFT spectral function and projected density of states of M$_2$ phase at 332 K, considering (a) antiferromagnetic and (b) paramagnetic states. In (a) and (b), the projections to $t_{2g}$ states associated with dimerized and zigzag-like chains of V-atoms are shown in the central and right panels, respectively. In respect to the dimerized atoms, projections to $a_{1g}$, $e_{g}^{\pi}(1)$, and $e_{g}^{\pi}(2)$ ``molecular'' states are shown in blue, red, and green, respectively. For the zigzag V-atoms, we use the same colors to each $t_{2g}$ state, with the contributions of the $up$ and $down$(dn) spins shown in solid and dashed lines, respectively. In (c) and (d) we show the valence histograms of the V-dimer and V states of dimerized and zigzag atoms, respectively.}
\label{fig:figM2awk_histo}
\end{figure*}

From these self-energies we notice a similar feature in both systems: the absence of poles in the imaginary part of the self-energies indicates that in bct phase, as in the M$_1$ phase, once the dimerization occurs, the Mott instability is arrested. In particular, we notice that the position of the peaks associated with the $a_{1g}$ antibonding states, which appear at -2.58 eV (bct) and -1.0 eV (M$_1$), indicates that the electrons in M$_1$ phase are closer to a Mott transition than the ones in bct phase. This suggests that the structural distortions are more important in the gap opening of bct phase than in the M$_1$ phase. 

In Fig.~\ref{fig:self_M1bct_onreal}(b) and (c) we show the valence histograms of vanadium and niobium dimers in M$_1$ and bct phases, respectively. These histograms indicate that the singlet states (blue arrows) associated with the V and Nb-dimers are the states with the highest probabilities, with occupation probabilities of $32\%$ and $35\%$, respectively, followed by impurity states with 3 and 1 electrons ($N=3$ and $1$ in our histograms). These findings suggest that charge fluctuations are more important to the gap formation than spin 
fluctuations associated with the singlet-triplet states of the dimers. In respect to this, we mention that recent inelastic X-ray scattering measurements found that the singlet-triplet spin excitation energy is 0.42 eV,~\cite{heRIXS} in disagreement with the theoretical prediction of 0.123 eV obtained by quantum Monte Carlo calculations.~\cite{WagnerQMC} 
The fact that probability for the triplet state is relatively small in our calculation, suggests that the singlet-triplet splitting is comparable to the band gap, but the precise calculation of the many body energy level is beyond the scope of this work.

\subsubsection{M$_2$ phase}
\label{res_M2}
Previous experimental works have shown that M$_2$ phase of VO$_2$ can be stabilized at ambient conditions by uniaxial stress along the $[110]_{R}$ axis or by doping with $3+$ ions, such as Cr$^{3+}$, Al$^{3+}$, Fe$^{3+}$, or Ga$^{3+}$.~\cite{pouget2,strelcov}
As can be seen in Fig.~\ref{fig:structures}(c), in this phase half of vanadium atoms dimerize, without tilting, whereas the other half experience a zigzag-like distortion along the $c$ axis. Further, previous experiments reported the existence of localized $d$ electrons in the zigzag vanadium chains. Pouget~\textit{et al.}~\cite{pouget1} interpreted their findings using a set of noninteracting independent spin-1/2 Heisenberg chains. Likewise, D'Haenens \textit{et al.}~\cite{merenda} deduced from their findings an antiferromagnetic exchange coupling $J_{AF}$, on the zigzag chain, of the order of 400 K, at $x = 0.45$ in V$_{1-x}$Cr$_{x}$O$_2$ compounds. More recently, such antiferromagnetic ordering has been observed in VO$_2$ nanorods in the M$_2$ phase.~\cite{jpark}

\begin{figure*}[!htb]
\includegraphics[scale=0.48]{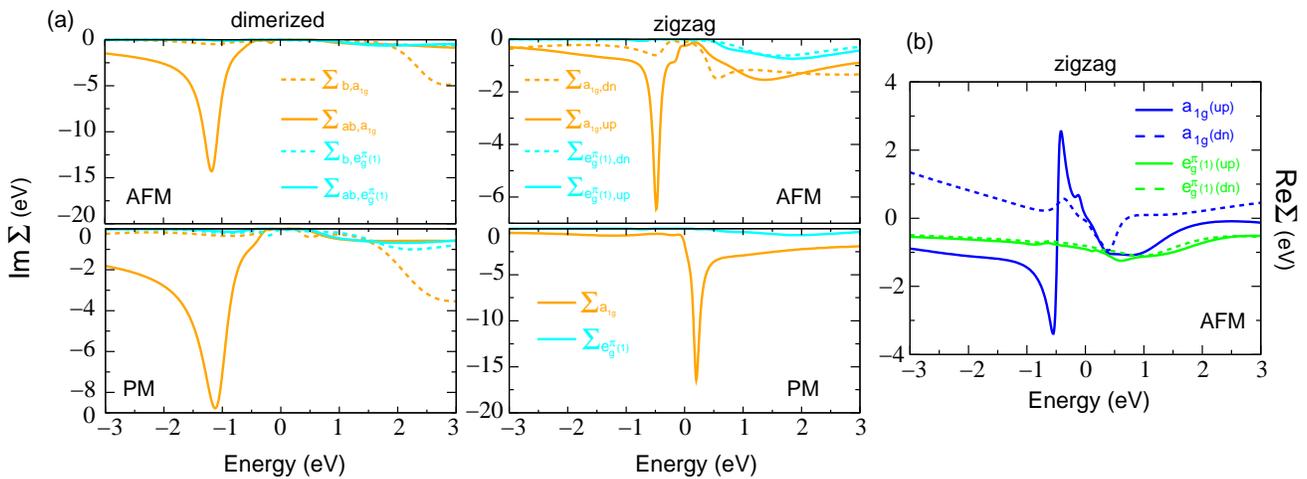}
\caption{(a) Imaginary part, on real frequency axis, of bonding and antibonding self-energies of $a_{1g}$ and $e_{g}^{\pi}(1)$ states of V-dimers (on the left) and of $a_{1g}$ and $e_{g}^{\pi}$ states of zigzag vanadium atoms (on the right). (b) Real part of $a_{1g}$ and $e_{g}^{\pi}$ self-energies of zigzag (unpaired) vanadium atoms in the antiferromagnetic phase.}
\label{fig:self_M2}
\end{figure*}

As reported previously,~\cite{whb} in our investigation of the M$_2$ phase we considered paramagnetic and antiferromagnetic states concerning the zigzag V-atoms.
From our calculated spectral functions, shown in Fig.~\ref{fig:figM2awk_histo}(a) and (b), we observe that in both states the M$_2$ phase presents a gap. In particular, we obtain gaps of 0.58 and 0.61 eV for the antiferromagnetic and paramagnetic states, respectively. This indicates that the antiferromagnetic ordering related to the $d$ electrons in the zigzag-like chains plays a minor role in the gap formation of this phase. In both situations the gap opens between the $a_{1g}$ and $e_{g}^{\pi}$ subband, although the $a_{1g}$ states from the zigzag chains provide a small contribution to the bottom of the conduction band (see central and right panels of Fig.~\ref{fig:figM2awk_histo}(a) and (b)).
The $a_{1g}$ bonding subband presents a coherent peak at -0.34 eV (-0.31 eV), whereas the antibonding subband is centered at 2.5 eV (2.5 eV) for the paramagnetic (antiferromagnetic) state. The resulting bonding-antibonding splitting energy is $\approx$ 1 eV larger than obtained by DFT. 
Further, the $a_{1g}$ states of zigzag atoms downshift and become less dispersive in the paramagnetic phase, such as a Hubbard-like subband. The $a_{1g}$ states of dimerized V-atoms do not shift. Hence, the low-energy excitations in the paramagnetic phase are dominated by the $a_{1g}$ states of the dimerized V-atoms, in contrast to the antiferromagnetic phase wherein both vanadium chains contribute to the top of the valence band.

We also observe a restoration of coherence of the occupied states close to the Fermi level when comparing the antiferromagnetic with the paramagnetic state, which indicates that antiferromagnetic ordering suppresses the electronic correlations. 
In fact, in the valence histogram shown in Fig.~\ref{fig:figM2awk_histo}(c) and (d), we observe that the antiferromagnetic ordering suppresses the spin fluctuations of the zigzag V-atoms, while it does not affect the occupation probabilities of states associated with the dimerized atoms. The histogram shown in Fig.~\ref{fig:figM2awk_histo}(c) indicates that the singlet state associated with the V-dimers has the highest probability (occupation probability of $\approx 33\%$), followed by states with $N=3$, as found for the M$_1$ phase.

To investigate the effects of electronic correlations in the M$_2$ phase we show in Fig.~\ref{fig:self_M2}(a) the imaginary part of self-energies related to the dimerized and zigzag V-atoms. In the paramagnetic phase, we notice that the $t_{2g}$ states associated with the dimerized atoms do not present any pole in the imaginary part of self-energy, as similarly found in M$_1$ phase. As pointed out in our previous work,~\cite{whb} in the antiferromagnetic phase we find that even the $t_{2g}$ states associated with the zigzag atoms do not present a Mott instability. In fact, the singularity of the self-energy is arrested once the antiferromagnetic ordered state is stabilized. Interestingly, as can be seen in Fig.~\ref{fig:self_M2}(b), the real part of $a_{1g}$ self-energy has a strong frequency dependence around the Fermi level, which indicates that the $a_{1g}$ subband is renormalized by this component.
On the other hand, in the paramagnetic phase (see Fig.~\ref{fig:self_M2}(a)), the imaginary part of self-energy associated with $a_{1g}$ states of zigzag V-atoms acquires a pole. As a result, the $a_{1g}$ subband is splitted by a Mott instability, indicating that this subband undergoes the Mott-Hubbard transition. These findings suggest that the M$_2$ phase is best characterized as a Mott insulator.

\subsubsection{Nonlocal dynamic correlations in M$_1$, M$_2$, and bct phases}
\label{res_nonloc}

\begin{figure}[!htb]
\includegraphics[scale=0.6]{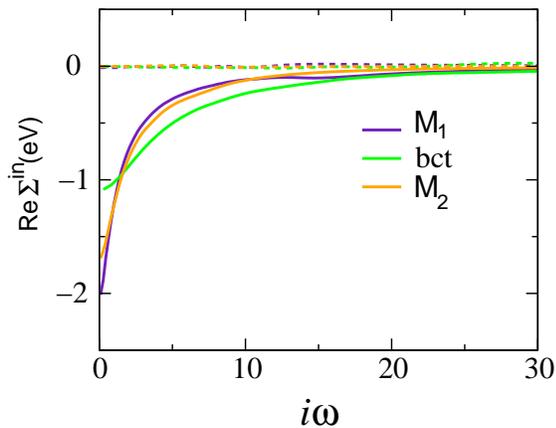}
\caption{Real part of intersite self-energies, on imaginary frequency axis, of $a_{1g}-a_{1g}$ (solid lines) and $e_{g}^{\pi}(1)-e_{g}^{\pi}(1)$ (dashed lines) states of M$_1$ (indigo), M$_2$ (AFM) (orange), and bct (green) phases. We considered $T = 332$~K for the monoclinic phases and $T = 1000$~K in the case of bct phase.}
\label{fig:self_inter_Mbct} 
\end{figure}

As observed in the previous sections, the inclusion of nonlocal dynamic correlations increases the $a_{1g}$ bonding-antibonding splitting energy in the low-temperature phases of VO$_2$ and NbO$_2$. Within the transition metal dimers, treated as a cluster in our DMFT calculations, it is useful to look at the self-energies in the site representation, where we have the local self-energy, $\Sigma^{local} = \Sigma_{11}$, and the intersite self-energy, $\Sigma^{in} = \Sigma_{12}$. 
In order to compare the effects of nonlocal dynamic correlations for the M$_1$, M$_2$, and bct phases, we show in Fig.~\ref{fig:self_inter_Mbct} the real part of intersite $a_{1g}-a_{1g}$ and $e_{g}^{\pi}(1)-e_{g}^{\pi}(1)$ self-energies for each insulating phase.

First, we notice that the frequency dependence of the intersite self-energies associated with $e_{g}^{\pi}(1)$ states is negligible for all the insulating phases. In contrast, the intersite self-energies associated with $a_{1g}$ states depend strongly on frequency in the low-energy part. This indicates the presence of strong intersite correlations within the transition metal dimers. This strong frequency dependence of intersite self-energy was first noticed in the low-temperature phase of Ti$_2$O$_3$,~\cite{poteryaev} but a strong intersite Coulomb interaction was required for opening the gap.
In the M$_1$ and M$_2$ phases of VO$_2$ the intersite components are almost the same, with minor difference in the $i\omega \rightarrow 0$ limit, indicating that $|\Sigma^{in}_{a_{1g}-a_{1g}}(0)|$ is larger in M$_1$ phase. Interestingly, the intersite $a_{1g}$ self-energy in bct phase is frequency dependent as well, but with smaller intensity than in the insulating phases of VO$_2$, as noticeable when taking the $i\omega \rightarrow 0$ limit. These results, within an effective band structure picture (see supplemental material of Ref.~\onlinecite{whb}), support the smaller increase of bonding-antibonding 
splitting energy in bct phase in comparison with the monoclinic phases of VO$_2$. Therefore, our findings suggest that the Mott physics is important in both oxides, with only somewhat stronger effects in VO$_2$.

\section{Summary}
\label{conclu}
In summary, we performed fully self-consistent all-electron DFT+DMFT calculations to investigate the role of dynamic electronic correlations on the electronic structure of the metallic and insulating phases of NbO$_2$ and VO$_2$, as well as the mechanism of the gap opening for the insulating phases, focusing on a comparison between the two compounds.
For the rutile phase of both oxides our results indicate that dynamic correlations lead to a renormalization of $t_{2g}$ levels and the emergence of Hubbard bands associated with the $a_{1g}$ (LHB) and $e_{g}^{\pi}$ (UHB) states. In particular, we find that the correlation effects are more pronounced in the spectral properties of VO$_2$, although the calculated quasiparticle weights $Z$'s show that electronic dynamic correlations in the rutile phase of NbO$_2$ are still important. The smaller values of $Z$ obtained for VO$_2$ reveal that electrons in this system are closer to the Mott transition than electrons in NbO$_2$, which is expected due to more delocalized nature of $4d$ orbitals of niobium in comparison with the $3d$ ones of vanadium atoms. 

In respect to the insulating phases of both oxides, we find charge gaps of 0.55 eV and 0.73 eV for the M$_1$ and bct phases, respectively, in agreement with experiments. 
For the M$_2$ phase, by its turn,  we obtain charge gaps of 0.58 eV, considering an antiferromagnetic state, and of 0.61 eV, for the paramagnetic state. This indicates that antiferromagnetic ordering plays a minor role in the gap formation of this phase. Overall, we observe that the bonding-antibonding splitting energy increases in the presence of nonlocal dynamic correlations, in comparison with our DFT calculations.
Interestingly, we find that nonlocal dynamic correlations do play a role in the gap formation of the bct phase, by a similar physical mechanism as in the case of M$_1$ phase of VO$_2$ which was proposed in Ref.~\onlinecite{whb}. In particular, the nonlocal dynamic correlations in bct phase are less important for the gap opening than in M$_1$ and M$_2$ phases of VO$_2$. It indicates that the bct phase of NbO$_2$ is not a purely Peierls-type insulator, as it was recently proposed.~\cite{oharaPRB} According to our results, all phases of VO$_2$ and NbO$_2$ are in the near vicinity of a Mott transition, but with the Mott instability arrested in the dimerized and antiferromagnetic phases.

\section{Acknowledgments}
We acknowledge support from the Brazilian agencies CNPq, CAPES, and FAPEMIG. K.H. and G.K. were supported by NSF DMR-1405303 and NSF DMR-1308141, respectively.

\end{document}